\documentclass[twocolumn]{aastex631}

\begin{document}

\title{Emergence of cHz Quasi-Periodic Oscillations from low angular momentum flow onto a supermassive black hole}

\correspondingauthor{Indu K. Dihingia, Yosuke Mizuno}
\email{ikd4638@sjtu.edu.cn, ikd4638@gmail.com, mizuno@sjtu.edu.cn}

\author[0000-0002-4064-0446]{Indu K. Dihingia}
\affiliation{Tsung-Dao Lee Institute, Shanghai Jiao-Tong University, 1 Lisuo Road, Shanghai, 201210, People's Republic of China}

\author[0000-0002-8131-6730]{Yosuke Mizuno}
\affiliation{Tsung-Dao Lee Institute, Shanghai Jiao-Tong University, 1 Lisuo Road, Shanghai, 201210, People's Republic of China}
\affiliation{School of Physics \& Astronomy, Shanghai Jiao-Tong University, 800 Dongchuan Road, Shanghai, 200240, People's Republic of China}
\affiliation{Institut f\"{u}r Theoretische Physik, Goethe Universit\"{a}t, Max-von-Laue-Str. 1, 60438 Frankfurt am Main, Germany}
\affiliation{Key Laboratory for Particle Physics, Astrophysics and Cosmology, Shanghai Key Laboratory for Particle Physics and Cosmology, Shanghai Jiao-Tong University,800 Dongchuan Road, Shanghai, 200240, People's Republic of China}
 
\begin{abstract}
Quasi-periodic oscillations (QPOs) are very common in black hole accretion systems that are seen from the modulations in luminosity. Many supermassive black hole sources (e.g., RE J1034+396, 1H~0707-495, MCG-6-30-15, 1ES~1927+654, Sgr~A$^*$) have been observed to exhibit QPO-like variability in the range of mHz in different energy bands (e.g., radio, NIR, X-ray). Due to the shorter infalling time, low-angular momentum accretion flows can have resonance close to the black hole, which will raise variability centiHz (cHz) or beyond QPOs for supermassive black holes. In this study, we for the first time show that such resonance conditions can be achieved in simulations of low angular momentum accretion flows onto a black hole. The QPOs could have values beyond $\nu_{\rm QPO}\gtrsim0.1-1\times 10^7M_\odot/M_{\rm BH}\,$cHz and the harmonics have a ratio of 2:1. Hunting down of these cHz QPOs will provide a smoking gun signature for the presence of low-angular momentum accretion flows around black holes (e.g., Sgr~A$^*$, 1ES~1927+654).
\end{abstract}


\section{Introduction}\label{section1}

Variability in supermassive black holes (SMBHs) across many wavelengths is a hallmark signature of active galactic nuclei (AGN) \citep{Elvis-etal1994,Hovatta-etal2007,Ricci-etal2017}. Such variability may be due to jet activity, disk instabilities, and/or accretion turbulence. 
M87, with its large-scale relativistic jet, shows radio and gamma-ray variability on days to months \citep{Hada-etal2014,Satapathy-etal2022}. Sgr~A$^*$, the quiescent low luminous SMBH at the center of our galaxy, has flares in the infrared and X-ray bands that happen every few minutes to hours \citep{Witzel-etal2018,GRAVITY2020}. The theoretical understanding of it is still elusive (for discussion, check ~\citep{Olivares-etal2023,Dihingia-Mizuno2024}). On the other hand, 1ES~1927+654, which transitioned from a type-II to type-I AGN within months in 2018 due to a sudden accretion disk disruption \cite{Trakhtenbrot-etal2019}. Notably, it shows minute-scale variability due to processes happening in the innermost part of the accretion disk \citep{Ricci-etal2020,Masterson-etal2025,Laha-etal2025}.

Variabilities are often observed in the quasi-periodic oscillations (QPOs) in power density spectra (PDS). Sometimes, SMBH sources show extremely short time scales (minutes to hours) in their light curves in the mHz to sub-mHz range (for example, RE~J1034+396 \citep{Gierlinski-etal2008}, 1H~0707-495 \citep{Pan-etal2016}, MCG-6-30-15 \citep{Vaughan-etal2005}, 1ES~1927+654 \citep{Ricci-etal2020,Masterson-etal2025,Laha-etal2025}, Sgr~A$^*$ \citep{Iwata-etal2017,Haggard-etal2019,Iwata-etal2020}). These time scales are analogous to high-frequency QPOs (HFQPOs) in black hole X-ray binaries (BH-XRBs), considering the stellar-mass black hole instead of SMBH due to mass scalling \cite[see][]{McHardy-etal2006}.
Although we do not have clear evidence of the physical origin of QPOs, several mechanisms are proposed. Oscillations of transition layers (the boundary between Keplerian and sub-Keplerian matter), coronae, standing shocks \citep[e.g.,][]{Titarchuk-Fiorito2004,Cabanac-etal2010,Das-etal2014}, accretion-ejection instability \citep[e.g.,][]{Tagger-Pellat1999,Varniere-etal2002}, or relativistic precession model \cite[e.g.,][]{Stella-Vietri1998,Schnittman-etal2006,Ingram-eta2009} could explain low-frequency QPOs (LFQPOs) or lower ranges of HFQPOs. For upper ranges of HFQPOs, the relativistic precession model, the warped disk model, and resonance models are proposed, \cite[e.g.,][]{Aliev-Galtsov1981,Stella-Vietri1998,Abramowicz-Kluzniak2004,Kato2004}. ``p-mode'' oscillations of acoustic waves within a relatively smaller accretion torus could also explain such HFQPOs \cite{Rezzolla-etal2003b}.

The weak magnetic fields in accretion flows are potentially unstable to the magneto-rotational instabilities (MRI) \citep{Balbus-Hawley1991,Balbus-Hawley1998}.
The inherent turbulence in accretion flows driven by MRI can grow if the accretion timescale is longer than the turbulence timescale which leads to the transport of angular momentum and accretion onto a black hole. With the decrease of angular momentum in accretion flows, they are not in equilibrium with the gravity of the black hole. The flows directly fall onto a black hole without angular momentum transport by turbulence. 
Thus the accretion time scale becomes shorter. When the accretion time scale is shorter than the turbulent one, the turbulence cannot grow anymore and saturate. This gives us opportunities to harness unsaturated turbulence in the time series. Depending on the infalling time, different timescales (infalling and turbulence) may satisfy the resonance condition, giving rise to QPO-like oscillation features in accretion flows. In this study, we for the first time demonstrate it with new sets of 3D GRMHD simulations of low angular momentum accretion flows onto the rotating Kerr black hole. 
These QPOs may range from the cHz range and beyond for SMBHs. It is suggested that the matter with low angular momentum may explain some of the complex behavior around SMBH Sgr~A$^*$ at the galactic center \citep{Ressler-etal2018,Dihingia-Mizuno2024}. 
Our study provides unique observational possibilities of low-angular momentum flows around SMBHs in terms of very rapid variabilities.

\section{Numerical methods}

\subsection{General relativistic magneto-hydrodynamic simulations} 

GRMHD has been the favorite as well as the essential theoretical tool of the last decades to study accretion around black holes, especially after the first direct images of M\,87$^*$ and Sgr\,A$^*$  and their reconstructions based on it\cite{EHTC2019,EHTC2022}.
This study examines low angular momentum accretion flows using 3D ideal GRMHD simulations with the GRMHD code \texttt{BHAC}\cite{Porth-etal2017, Olivares-etal2019} in Modified Kerr-Schild coordinates. We use spherical polar grids $(r, \theta, \phi)$ with logarithmic spacing in the radial direction ($r$, up to $r=2500\,r_g$) and linear spacing in the polar and azimuth directions. Simulations are performed in a generalized unit system with $G=M_{\rm BH}=c=1$. Here, $G$, $M_{\rm BH}$, and $c$ represent the universal gravitational constant, black hole mass, and light speed, respectively. Length and time scales are expressed in units of $r_g=GM_{\rm BH}/c^2$ and $t_g=GM_{\rm BH}/c^3$. To make the efficient use of our numerical resources, we performed all simulations in lower resolutions ($256\times80\times64$). Due to low angular momentum, the accretion process takes place automatically without transporting angular momentum outwards by magneto-rotational instabilities (MRI). However, to check consistency, we also perform one simulation with higher resolution ($512\times160\times128$, with two static mesh refinement levels), where higher resolution is concentrated within the region $\pm65^\circ$ from the equatorial plan and $r\le100\,r_g$. This is reasonable since the physics we intend to study happens very close to the black hole.

We set the initial density using the flattened Fishbone-Moncrief (FM) torus \cite{Fishbone-Moncrief1976} distribution 
$\rho (r,\theta)=\rho_{\rm FM}(r,\theta) \exp(-\alpha^2 \cos\theta^2).$
Here, $\rho_{\rm FM}(r,\theta)$ is the density distribution described by rotationally-supported torus around a rotating black hole \cite{Fishbone-Moncrief1976}. To obtain a low-angular momentum flow solution with this setup, we set the inner edge at $r_{\rm min}=6\,r_g$ and the maximum density at $r_{\rm max}=15\,r_g$ following \cite{Dihingia-Mizuno2024}. Here, the value of $\alpha (=10)$ decides the altitude of the inflow, such choice provides us with a flattened disk-like distribution of matter around the equatorial plane. The impacts on the flow structure and the PDS of the accretion rate are shown in \ref{Appendix-B}. For low-angular momentum flows, we supply an angular momentum ($\lambda_0$) of the flow as a fraction of Keplerian angular momentum (${\cal F}$) at the position of $r_{\rm max}$, where $\lambda_0$ is the maximum angular momentum of the flow (follow Refs. \cite{Dihingia-Mizuno2024} for detail). Additionally, we supply a non-zero azimuthal component of the vector potential $A_\phi \propto \max(\rho/\rho_{\rm max}-10^{-3},0)$. We chose the strength of the magnetic field by setting the minimum value of the initial plasma-$\beta$ to be $\beta_{\rm min}=100$ for all the simulation models. The details of all simulation models are listed in table~\ref{tab-01}. In table~\ref{tab-01}, we display the value of $\lambda_0$ in terms of percent of marginally stable angular momentum at the fifth column, i.e., $\lambda_{\rm ms}$. Additionally, we show the total simulation times for different models, the Kerr parameter ($a$), and the resolution for different models at the 6th to 8th columns, respectively. 
%*************************************

\begin{table*}[t]
\centering
\caption{Model parameter. The explicit values of angular momentum fraction ${\cal F}$, specific angular momentum ($\lambda_0$), corresponding percentage of $\lambda_{\rm ms}$, simulation time, Kerr parameter, and resolution for different models are displayed.}
  \begin{tabular}{| c | l | c | c | c | c | c | c | c |}
    \hline
    Sl. No. & Model & ${\cal F}$ &$\lambda_0$ & $\%$ $\lambda_{\rm ms}$ & Sim. time ($t_g$) & $a$ & Resolution\\ 
    \hline
    1 & \texttt{MOD1} &  $0.10$ & $0.46$ & 19.3 & 5000 & 0.94 & $256\times80\times64$\\
    2 & \texttt{MOD2} &  $0.294$ & $1.378$ & 56.8 & 5000 & 0.94  & $256\times80\times64$\\
    3 & \texttt{MOD3} &  $0.30$ & $1.378$ & 58.0 & 5000 & 0.94  & $256\times80\times64$\\
    4 & \texttt{MOD4} &  $0.30$ & $1.378$ & 58.0 & 5000 & 0.94  & $512\times160\times128$\\
    5 & \texttt{MOD5} &  $0.45$ & $2.067$ & 87.0 & 5000 & 0.94 &  $256\times80\times64$\\
    6 & \texttt{MOD6} &  $0.48$ & $2.205$ & 92.8 & 12000 & 0.94 &  $256\times80\times64$\\
    7 & \texttt{MOD7} &  $0.49$ & $2.251$ & 94.8 & 12000 & 0.94 &  $256\times80\times64$\\
    8 & \texttt{MOD8} &  $0.50$ & $2.297$ & 96.7 & 12000 & 0.94  & $256\times80\times64$\\
    9 & \texttt{MOD9} &  $0.30$ & $1.379$ & 56.7 & 7000 & 0.92  & $256\times80\times64$\\
   10 & \texttt{MOD10} &  $0.306$ & $1.407$ & 57.9 & 5000 & 0.92  & $256\times80\times64$\\
   11 & \texttt{MOD11} &  $0.312$ & $1.434$ & 59.0 & 5000 & 0.92  & $256\times80\times64$\\
   12 & \texttt{MOD12} &  $0.318$ & $1.462$ & 60.1 & 5000 & 0.92  & $256\times80\times64$\\
    \hline
  \end{tabular}
\label{tab-01}
\end{table*}

\subsection{General Relativistic Radiative Transfer} 

GRRT calculation uses code \texttt{RAPTOR} \cite{Bronzwaer-etal2018,Bronzwaer-etal2020} to calculate the light curves at $230\,$GHz. \texttt{RAPTOR} solves polarized radiative transfer equations in curved spacetimes by ray-tracing method. We obtain the Stokes parameters ($I, Q, U, V$). Here we consider synchrotron radiation with thermal electron distribution function. 
In this study, we set the parameters for the GRRT calculations considering $M\,87^*$. Accordingly, we set the mass, distance, and inclination angle of the source to be $M_{\rm BH}=6.5\times10^9M_\odot$, $16.8\times10^3$ kpc, and $i=163$ degree, respectively. We adjust the mass scale by changing the mass accretion rate to obtain total intensity $\sim0.3\,Jy$.   Subsequently, we set the resolution $400 \times 400$ pixels of the images with $\pm 60\,r_{g}$ widths of field of view. We also calculate the electron temperature following the $R-\beta$ relation, i.e., $T_g/T_e = (R_{\rm low} + R_{\rm high}\beta^2)/(1 + \beta^2)$, where $T_g$, $T_e$, and $\beta$ correspond to gas temperature, electron temperature, and plasma-$\beta$, respectively \cite{Moscibrodzka-etal2016}. $R_{\rm low}=1$ and $R_{\rm high}=10$ are the constants that decide the maximum and minimum electron temperature, we fix them for this case study. To avoid the emission from the highly magnetized polar regions that affect the density floor treatment of GRMHD simulations, we set a cutoff value of magnetization ($\sigma=b^2/\rho$), $\sigma_{\rm cut}=1$ e.g., \cite{EHTC2019}, beyond which we neglect all the emission. Additionally, we compute emission only from close to the black hole ($r \le 100\,r_g$). We note that the emission from the outside region is negligible. 
%*****************************Figure 1***********************************
\begin{figure*}
\centering
\includegraphics[width=0.45\textwidth]{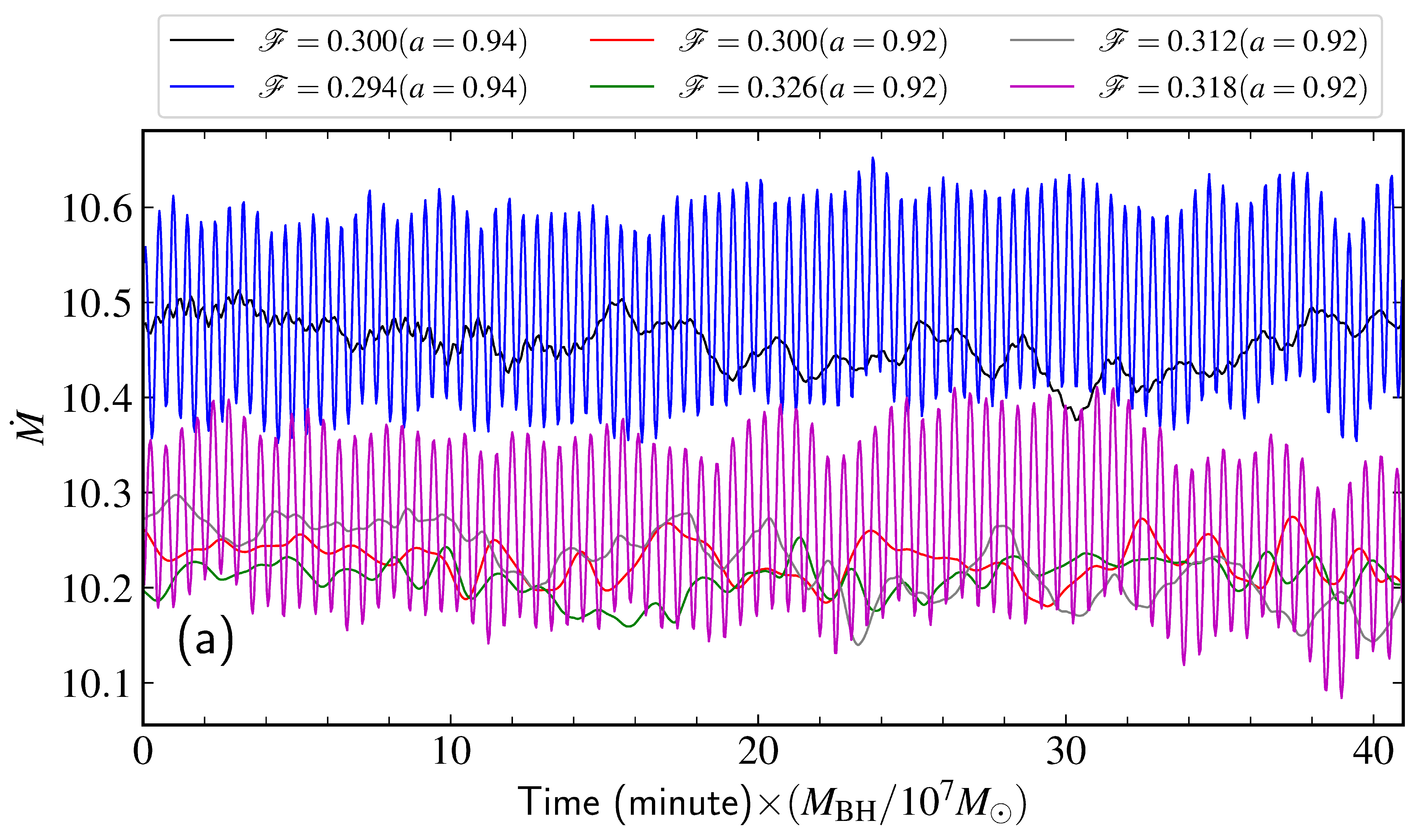}
\includegraphics[width=0.45\textwidth]{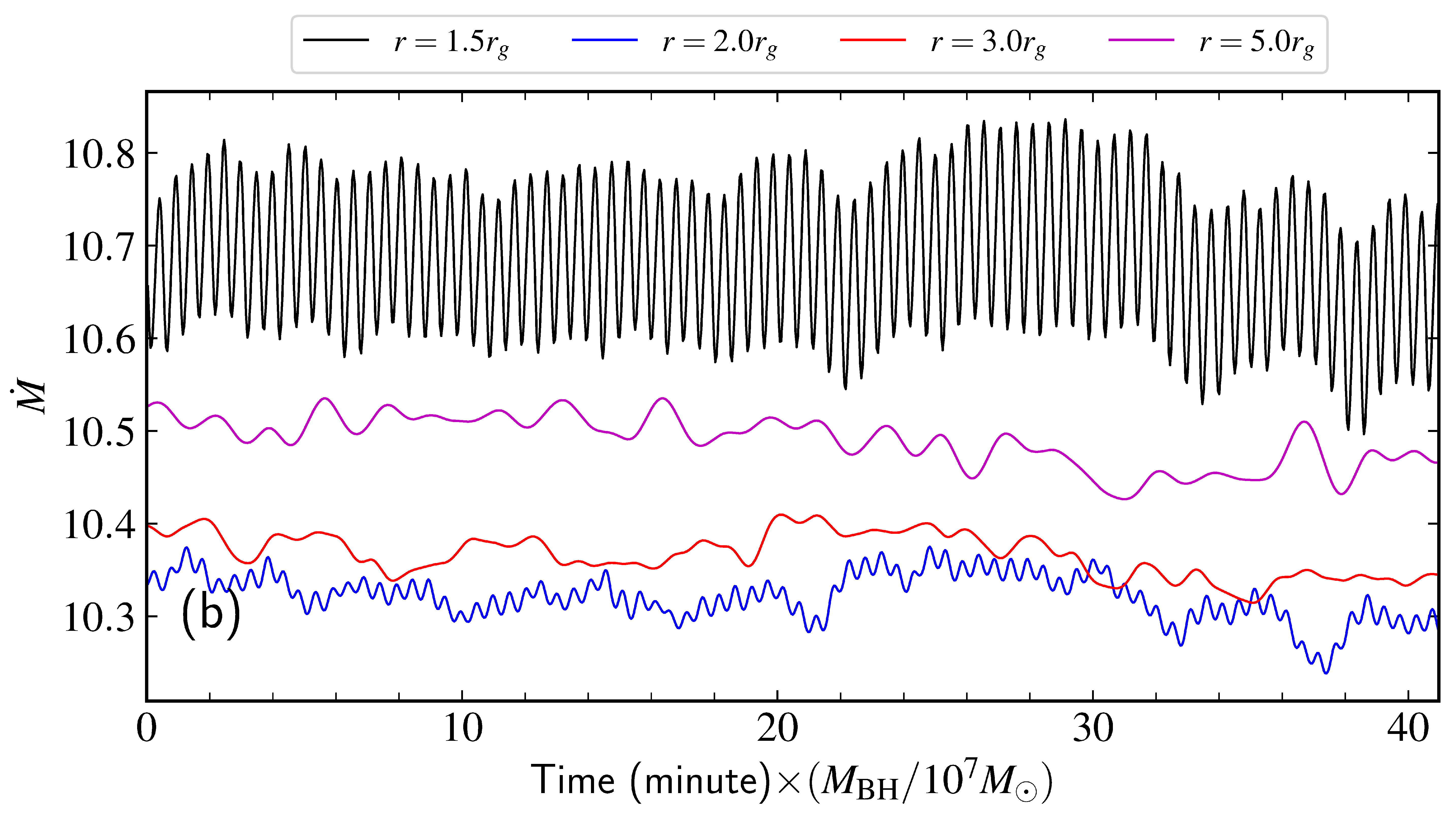}
\caption{Effect of tuning of angular momentum on the accretion rate profiles. {\it Left: (a)} Evolution of accretion rate at event horizon for two Kerr parameters $a=0.92$ and $a=0.94$ at different angular momentum around ${\cal F}=0.30$. {\it Right: (b)} Evolution of accretion rate at different radii ($r=1.5, 2.0, 3.0,$ and $5.0\,r_g$) for the cases of $a=0.92$ and ${\cal F}=0.318$. Different colors indicate different cases that are marked on the top of the panel.}
\label{Fig-01}
\end{figure*}
%*************************************************************************
%*****************************Figure 2***********************************
\begin{figure*}
\centering
\includegraphics[width=0.45\textwidth]{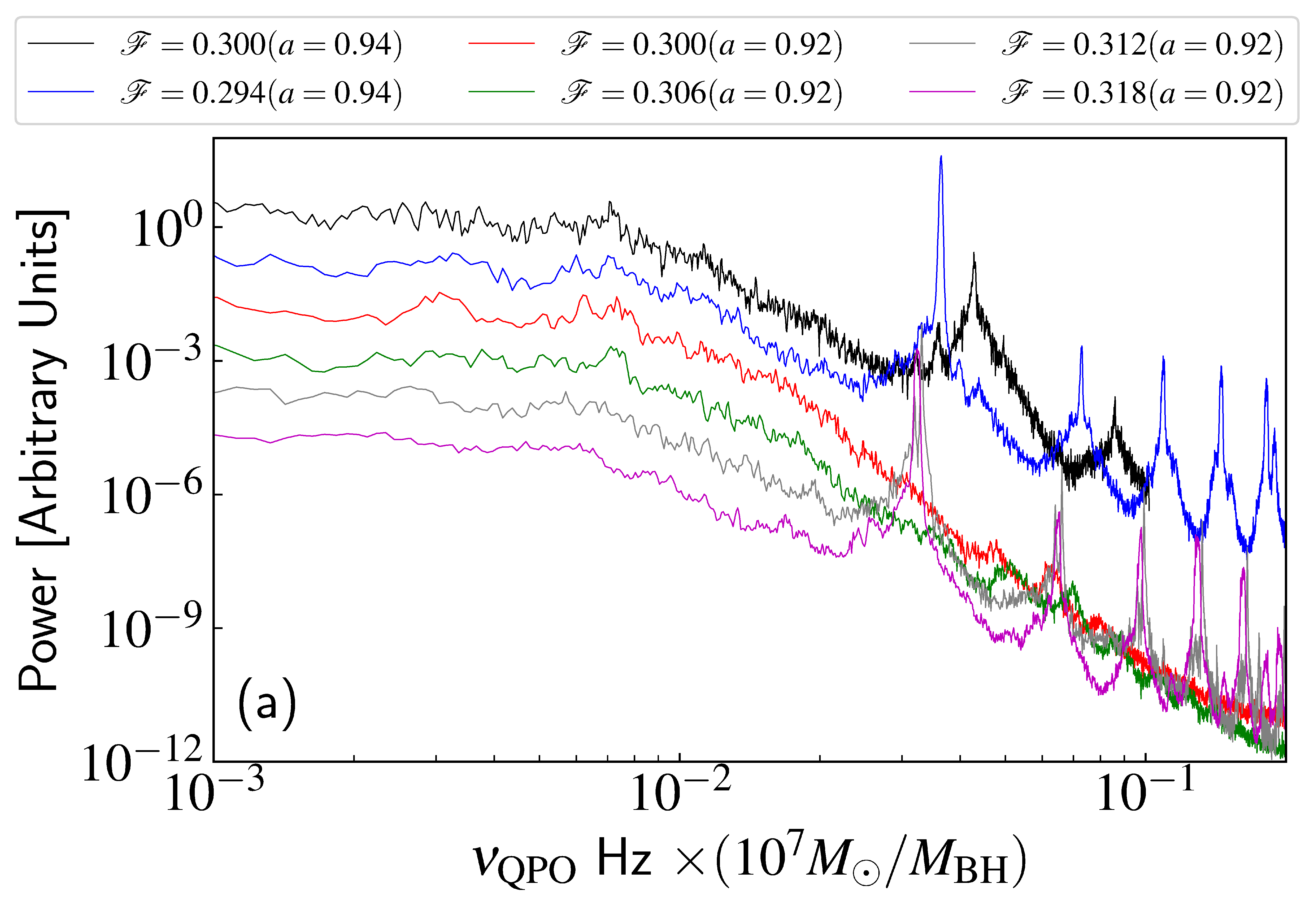}
\includegraphics[width=0.45\textwidth]{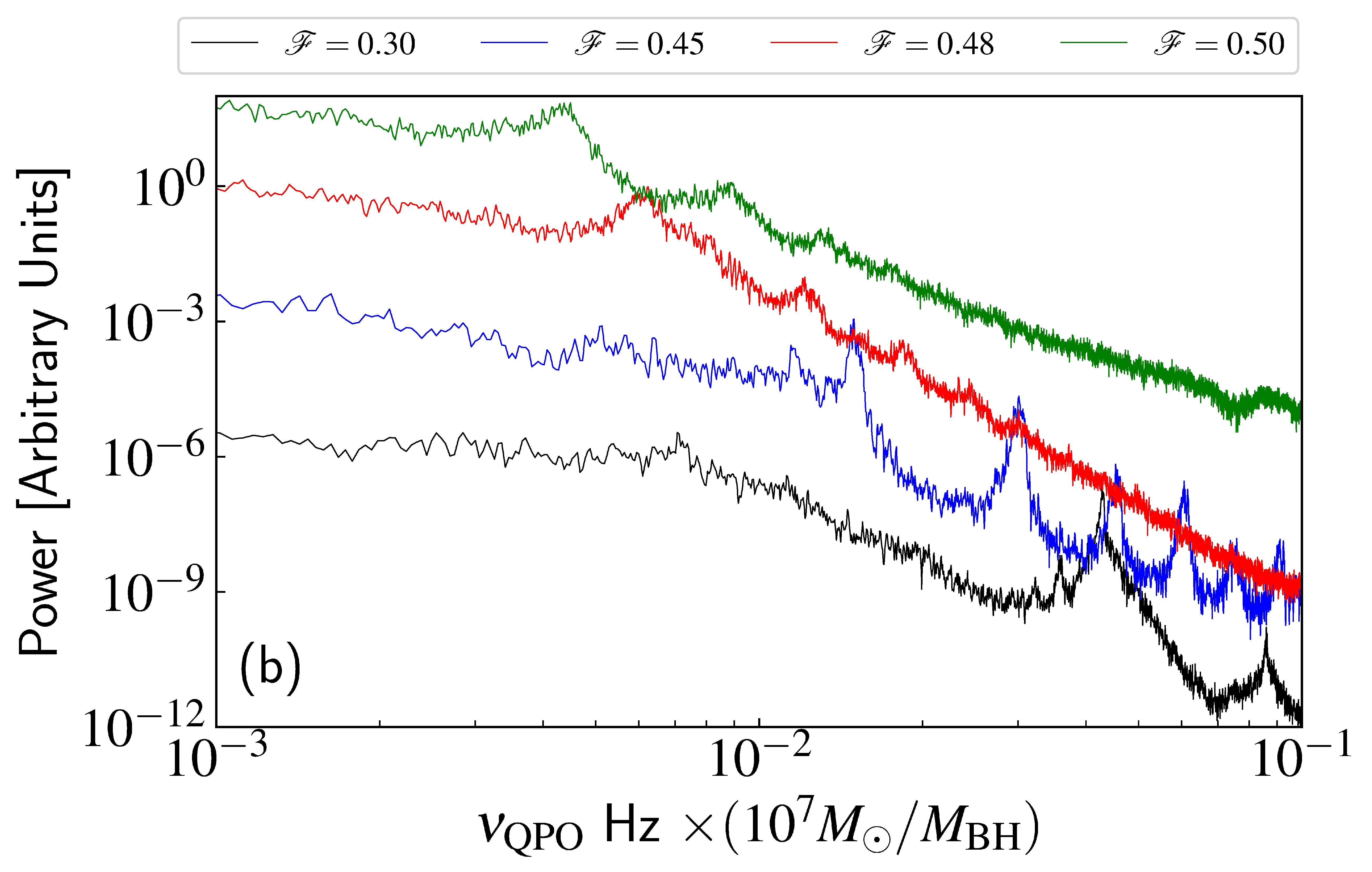}
\caption{Power density spectrum of accretion rate at the horizon in different angular momentum accretion flows. {\it Left: (a)} PDS of accretion rate at the event horizon for two different Kerr parameters $a=0.92$ and $a=0.94$ at different angular momentum cases around ${\cal F}=0.30$. {\it Right: (b)} Variation of PDS of accretion rate at the event horizon for the same Kerr parameter ($a=0.94$) cases with different angular momentum. Each value of angular momentum and Kerr parameters is marked on the top of the figure.}
\label{Fig-02}
\end{figure*}
%************************************************************************

\section{Near horizon accretion rate} \label{sec2}

If there is some resonance between the different timescales (e.g., infall time, MRI timescale, different epicyclic timescales), its signature is expected to appear on the accretion rate profiles. 
In this section, we extensively study it for different initial angular momentums and black hole spin parameters. In low-angular momentum flow, it is easy to vary the infall time by changing angular momentum with the parameter ${\cal F}$.  Fig.~\ref{Fig-01} shows the variation of accretion rate ($\dot{M}=\int\sqrt{-g}\rho u^r d\theta$ \cite{Porth-etal2017}) in code units at the horizon near the value ${\cal F}\sim 0.3$ for Kerr parameters $a=0.92$ and $a=0.94$. 
Considering our interest, we convert simulation times in minutes by scaling it for a SMBH with mass $M_{\rm BH}=10^7 M_\odot$.
For $a=0.94$, when we decrease angular momentum $2\%$ (${\cal F}=0.294$, blue curve) from the base value (black curve), the accretion rate profiles show very coherent oscillations, suggesting the presence of resonance.  Similarly, for $a=0.92$, we need to increase the value of ${\cal F}$ by $6\%$ (${\cal F}=0.318$, magenta curve) to get similar coherent oscillations. These oscillations can also be seen in different radii ($r=1.5$ and $2.0\,r_g$ ) close to the black hole, but the same oscillation is not visible in higher radii ($r=3.0$ and $5.0\,r_g$) (see Fig.~\ref{Fig-01}b). This suggests the resonance happens only close to the horizon, and accordingly, this can be affected by a strong gravity regime. These near-horizon modulations in the mass flux also influence the emission properties with the same modulations coming from this region (detail in Sec.~\ref{sec4}). Here, in Fig.~\ref{Fig-01}, we only show the accretion rate profiles from the range between $t=4950-5000\,t_g$ (in code units), a full version of the plots can be found in the \ref{Appendix-A}.

A more quantitative estimation of these oscillations is provided by calculating the power density spectra (PDS) of the accretion rate profiles. The oscillations appear as peaks in PDS. The coherent accretion rates correspond to very sharp peaks in the PDS; if they are not coherent, we do not find any sharp peaks or do not find any peaks. In Fig.~\ref{Fig-02}a, the variation of PDS is shown in the angular momentum values near ${\cal F}\sim0.3$ (same as Fig.~\ref{Fig-01}a). The PDS also presents harmonics clearly, i.e., multiple peaks in the same PDS. In this range of angular momentum, the first two prominent peak frequencies correspond to $\nu_{\rm QPO, p0}\simeq3.64\times10^{-2} (10^7M_\odot/M_{\rm BH})$Hz and $\nu_{\rm QPO, p1}\simeq7.28\times10^{-2} (10^7M_\odot/M_{\rm BH})$Hz, respectively, for Kerr parameter $a=0.94$.
We calculate the quality factors of these two peaks $Q=\nu_{\rm QPO,0}/\Delta\nu_{\rm QPO}$, which are given by $Q_{p0}\sim265$ and $Q_{p1}\sim226$, respectively. For the same Kerr parameter with ${\cal F}=0.3$, the quality factor of the primary peak frequency $\nu_{\rm QPO, p1}\simeq4.29\times10^{-2}\,(10^7M_\odot/M_{\rm BH})$~Hz is $Q_{p0}\sim140$. Thus, by tuning the angular momentum, the quality factor increases to higher values $Q\gg1$, suggesting the presence of resonance in the system. Similarly, for Kerr parameter $a=0.92$, the first two prominent peaks frequencies correspond to $\nu_{\rm QPO, p0}\simeq3.24\times10^{-2}\,(10^7M_\odot/M_{\rm BH})$~Hz and $\nu_{\rm QPO, p1}\simeq6.51\times10^{-2}\,(10^7M_\odot/M_{\rm BH})$~Hz, respectively. 
The quality factors for these two peaks are obtained as, $Q_{p0}\sim124$ and $Q_{p1}\sim142$, respectively. For both cases, the ratio between the peaks exactly corresponds to $2:1$.
Thus, frequencies have a range higher than $\nu_{\rm QPO}\gtrsim 30\,(10^7M_\odot/M_{\rm BH})$~mHz, which is much higher than any other characteristic frequencies. For reference, the maximum characteristic frequencies for $a=0.94$ and $M_{\rm BH}=10^7M_\odot$ at $r_{\rm ISCO}=1.904\,r_g$ can be easily calculated as: Keplerian frequency $\nu_{\rm K}\sim 2.97$~mHz, radial epicyclic frequency $\nu_r\sim2.1$~mHz, and vertical epicyclic frequency $\nu_\theta\sim2.97$~mHz. Similarly, for $a=0.92$, $r_{\rm ISCO}=2.17\,r_g$ and the characteristic frequencies are calculated as Keplerian frequency $\nu_{\rm K}\sim 2.58$~mHz, radial epicyclic frequency $\nu_r\sim1.85$~mHz, and vertical epicyclic frequency $\nu_\theta\sim2.58$~mHz, respectively.
Thus, the frequencies we observed here cannot correspond to any of these frequencies, and their origin must be the low-angular momentum flow itself. This characteristic radius ($r_{\rm ISCO}$) is not a property of low-angular momentum flow, and there is no known characteristic radius for such flow, which needs to be explored more in the future. Note that, in this work, we tuned the angular momentum for two individual spin cases and obtained coherent oscillations for both cases. Accordingly, we expect the same results can be achieved even in different spin parameters for the central black hole.

It is noted that for a given Kerr parameter, we will be able to get different peak frequencies at different angular momentum for the same Kerr parameter. However, to get very sharp peaks, one needs to adjust the angular momentum values appropriately. The details of this study are shown in Fig.~\ref{Fig-02}b, where the variation of PDS is displayed for different values of ${\cal F}$ for Kerr parameter $a=0.94$. For this study, we did not tune the values of ${\cal F}$ to get coherent oscillations. Still, we see a signature of the presence of oscillations in the PDSs. The first peaks frequency in the PDSs for ${\cal F}=0.30, 0.45, 0.48$, and $0.50$ corresponds to $\nu_{\rm QPO, p0}\equiv4.28, 1.49, 0.61,~{\rm and}~ 0.44\times10^{-2} (10^7M_\odot/M_{\rm BH})$ Hz, respectively.
The quality factors corresponding to these frequencies are obtained as: $Q_{p0}\sim247, 86, 14$, and $11$, respectively. In some of the cases, the quality factors are not very high ($\sim10$). For these cases, if angular momentum is tuned further, we can achieve higher $Q$ peaks as discussed in Fig.~\ref{Fig-02}a.
The frequency of the second peaks $(\nu_{\rm QPO, p1})$ in these PDSs also roughly follows $2:1$. 
We should note that the emergence of cHz QPOs in low-angular momentum flows is independent of numerical resolution; we confirmed it with higher resolution simulations (for details, see \ref{Appendix-C}). With higher resolution, the angular momentum needs to adjust properly to obtain the perfect resonance condition.

\section{Flow structure} \label{sec3}

%*****************************Figure 3***********************************
\begin{figure*}
\centering
\includegraphics[width=1\textwidth]{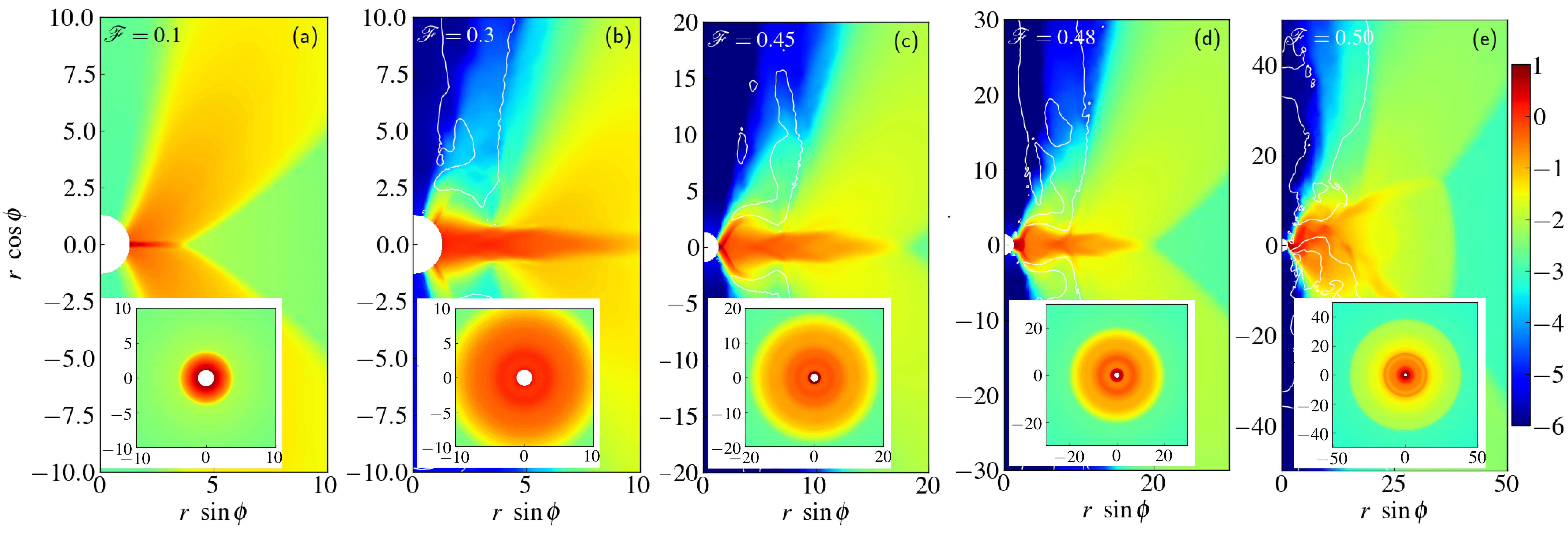}
\caption{Distribution of logarithmic density on the poloidal plane at time $t=5000\,t_g$. From left to right panels, different values of angular momentum (${\cal F}$) are used. The white contour corresponds to $u^r=0$ which is a boundary between inflow and outflow regions. In the inserted panels, it is shown the distribution of logarithmic density on the equatorial plane.}
\label{Fig-03}
\end{figure*}
%************************************************************************
%*****************************Figure 3***********************************
\begin{figure}
\centering
\includegraphics[width=0.45\textwidth]{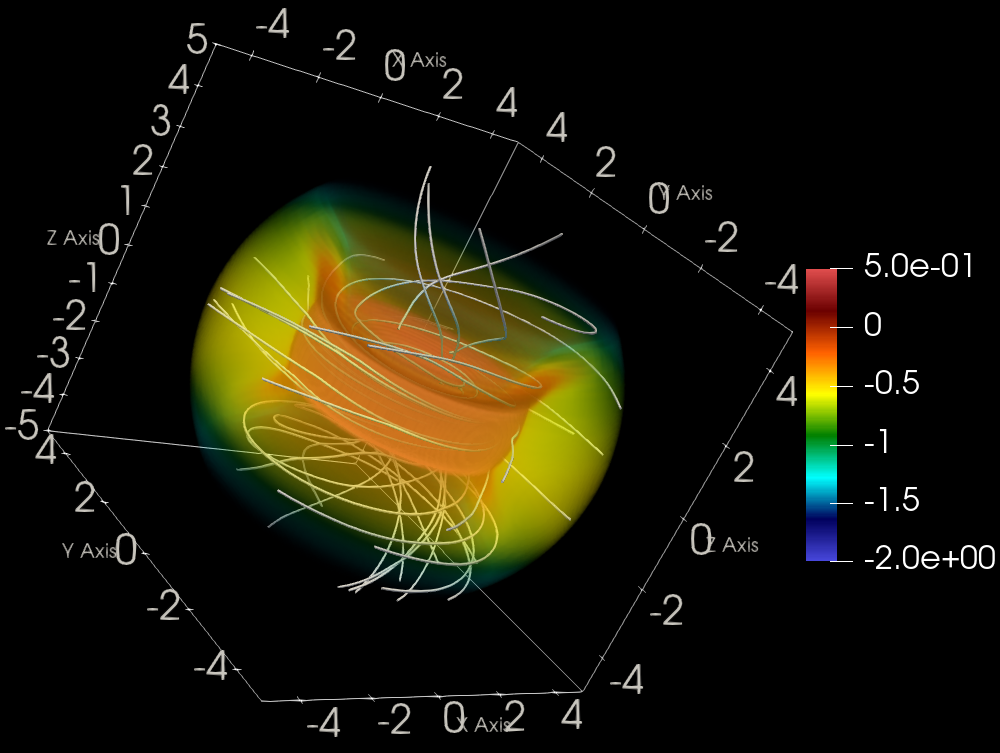}
\caption{3D volume rendering of logarithmic density distribution for Fig.~\ref{Fig-03}d to show formation PS close to the BH. Solid lines represent the magnetic field lines.}
\label{Fig-03a}
\end{figure}
%************************************************************************

The occurrence of cHz QPOs in low-angular momentum flows suggests the presence of some pseudo-surface (PS) close to the black hole. 
In low-angular momentum flow, when flow plunges onto the black hole, a small amount of matter tries to leave the system in a bipolar direction due to excess centrifugal force. It becomes stronger with the increase of angular momentum in accretion flows. The gradient of total pressure (gas and magnetic) will also contribute to launching the bipolar outflows.
If the angular momentum is enough to launch such an outflow, it creates a separation surface between inflow and outflow, which can act as a PS around a black hole. 
For a suitable angular momentum range, we see such formation of PS close to the black hole (see Fig.~\ref{Fig-03}). 
In different panels of Fig.~\ref{Fig-03}, we show the density distribution on the poloidal plane at $\phi=0$ for different values of angular momentum in accretion flows. In the insert, the same is shown on the equatorial plane. To mark the inflow-outflow surface, the radial 4-velocity $u^r=0$ contour is also displayed in all the panels. We do not see any outflow and PS in very low angular momentum cases (Fig.~\ref{Fig-03}a). This confirms the crucial roles of angular momentum in generating outflows and PS. In Fig.~\ref{Fig-03}b with ${\cal F} = 0.3$, we see the formation of a pseudo-surface (PS) close to the black hole (around $45^\circ$ degrees from the equatorial plane). With an increase in angular momentum (${\cal F} = 0.45$), the surface becomes more prominent and aligns to the equatorial plane (Fig.~\ref{Fig-03}c). With a further increase in the angular momentum, such as ${\cal F} = 0.48$, the surface moves far from the black hole (Fig.~\ref{Fig-03}d). 
While the infall time ($t_{\rm infall}$) and MRI growth timescale ($t_{\rm MRI}$) are approximate and depend on local disk conditions (e.g., density, magnetic field strength, and viscosity), their interplay can still lead to resonant-like behavior in a statistical or collective sense in certain suitable regions of the accretion flow. The approximate expression of the ratio of the timescales can be expressed as $t_{\rm MRI}/t_{\rm infall}\sim \alpha h^2/\sqrt{\beta}$ (following \cite{Frank-etal2002,Sano-etal2004}), where $\alpha$ and $h$, are the viscosity and aspect ratio of the flow, respectively. Low-angular momentum flow is usually geometrically thick, i.e., $h\sim1$. Therefore, in the region where $\alpha\sim\sqrt{\beta}$, we can have resonances between them. Such conditions may be satisfied in the boundary region of the inflow and outflow near PS. Once the resonance conditions are satisfied, this surface oscillates following the resonance frequency. To support our findings, in Fig.~\ref{Fig-03a}, we show a 3D volume rendering of the logarithmic density distribution of Fig.~\ref{Fig-03}d close to the black hole along with magnetic field lines. In the figure, we clearly see the formation of axisymmetric high-density PS. Towards the equatorial plane, we observe helically-twisted magnetic field lines due to inflow-dominated accretion, while in the bipolar region, we see more vertical magnetic field lines due to outflow.
We also observe a similar flow structure even in high-resolution simulations, confirming our results to be independent of resolution (for details, see \ref{Appendix-C}). Eventually, with a further increase in angular momentum (${\cal F} = 0.50$), we found sharp shock jumps seen in Fig.~\ref{Fig-03}e at a radius of $r\sim40\,r_g$. Such standing shocks usually modulate with respect to a mean position. The oscillation time scales of standing shock are much longer than the resonance of PS, \cite[e.g.,][]{Das-etal2014}. Thus, they could be associated with LFQPOs observed in black hole X-ray binaries. In order to see such oscillations in our simulations, we need to run them for a very long time $t\gg10^4\,t_g$. With current numerical resources, we cannot do such studies, and we plan to take them up in the future.
Note that targeting different astrophysical scenarios in semi-analytic, Newtonian, pseudo-Newtonian axisymmetric, and general relativistic hydrodynamic (GRHD) studies of shock solutions in the accretion flow around black holes have been widely studied, \citep[e.g.,][]{Molteni-etal1994, Proga-Begelman2003, Aktar-etal2015, Sukova-etal2017, Kim-etal2017, Dihingia-etal2018, Okuda-etal2019, Dihingia-etal2020}. Note that by increasing the angular momentum beyond the marginally stable value, we will have the flow behaviors studied earlier by various simulations using initial hydrostatic equilibrium torus
\cite[e.g.,][]{Porth-etal2017, Davis-Tchekhovskoy2020, Mizuno-etal2021,Dihingia-etal2023}.

%*****************************Figure 4***********************************
\begin{figure}
\centering
\includegraphics[width=0.45\textwidth]{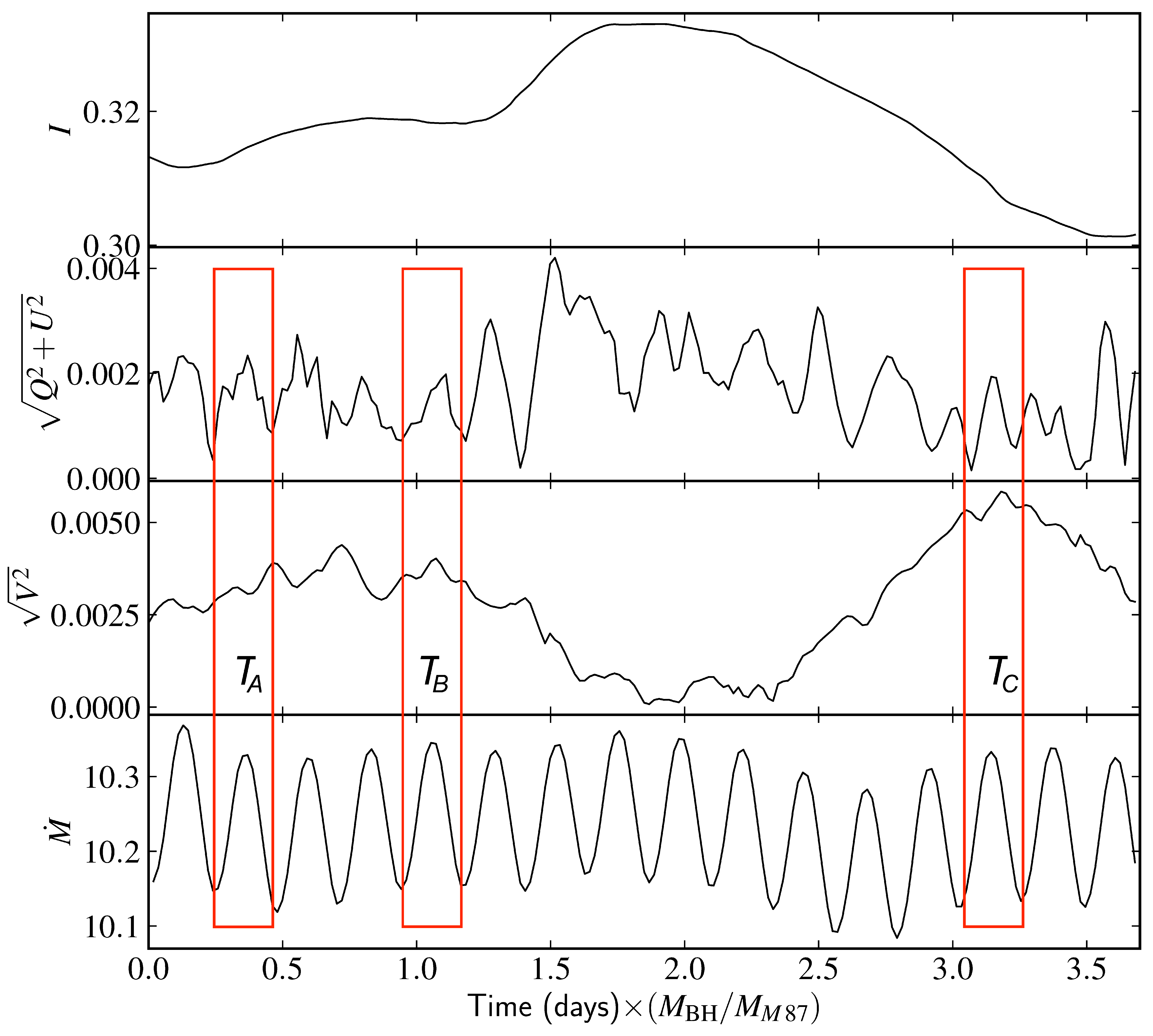}
\caption{Light-curve of polarized emission at $230\,$GHz. It is presented the case of ${\cal F}=0.318$ with $a=0.92$. Panels (a)-(c) show total intensity ($I$), linear polarization  ($\sqrt{Q^2 + U^2}$), and circular polarization ($\sqrt{V^2}$), respectively. Panel (d) presents the corresponding accretion rate ($\dot{M}$) at the horizon. Red rectangles show selected time-period of oscillations.}
\label{Fig-04}
\end{figure}
%************************************************************************

\section{Detectability of cHz QPOs} \label{sec4}

Although the accretion rate profiles show sharp peaks in their PDS, the variability magnitude of the accretion rate $\delta\dot{M}/\dot{M}\sim1\%$ at the horizon and far from the horizon it even drops to $\delta\dot{M}/\dot{M}\sim0.1\%$. Furthermore, the reported oscillation happens close to the event horizon. It is possible that the frequencies would not be in the very high-frequency range and would be red-shifted to zero for an observer at infinity.
Therefore, to check the observational detectability of these QPOs, we calculate the radiative signatures by using a general relativistic radiation transfer code. The radiations are calculated considering the mass of M\,87*, i.e., $6.5\times10^9\,M_{\odot}$\cite{EHTC2019}. 

Due to the magnetized nature of the accretion flows, the flow close to the black hole could emit synchrotron radiations. If some of these emissions leave the strong gravity and reach us with the QPO signature, it will possibly detect them. Accordingly, we show synchrotron emission characteristics in Fig.~\ref{Fig-04}. 
The units of time are shown in days considering M\,87 mass. The range of simulation time in code units corresponds to $t=4990-5000\,t_g$.
The total intensity ($I$) of the synchrotron emission at $230\,$GHz does not follow the accretion oscillations that are seen Fig.~\ref{Fig-04}a and \ref{Fig-04}d. This indicates that oscillations that happen near the horizon fade away to a distant observer due to gravitational red-shift.
However, these oscillations flicker the magnetic field structure close to the black hole. These flickering impacts the linear polarization patterns, as shown in Fig.~\ref{Fig-04}b. The peaks in the linear polarized emission ($\sqrt{Q^2 + U^2}$) and the accretion rate roughly follow a similar time evolution (compare with Fig.~\ref{Fig-04}d). In the figure, three time periods ($T_A, T_B,$ and $T_C$) are marked to show their correlations. On the contrary, the circular polarization ($\sqrt{V^2}$) does not show oscillations similar to the accretion rate. Oscillatory signatures are washed out for total intensity and circular polarization, which are undetectable. On the contrary, this analysis suggests that the high-resolution polarization observations will provide evidence of such high-frequency QPOs.

\section{Summary}

We found that low-angular momentum flow around black holes provides unique observational evidence. The resonances happening very close to the black hole make the PS oscillate, and they are reflected in the accretion rate profiles. 
A perfect resonance condition can be achieved by tuning the angular momentum. Due to the very short infall time of low angular momentum flow, the resonance happens at a very high frequency, which is about a few to a few $10$s of $(10^7M_\odot/M_{\rm BH})\,$~mHz. The prominent peaks in the PDS show harmonics with a ratio of $2:1$. Due to the weak strengths of the oscillations, their signatures can only be detected in the linearly polarized emission. 

These cHz QPOs could be very interesting in terms of gravity physics, as their possible origin is very close to the black hole event horizon. We should note that Sgr~A$^*$ shows QPO features of variations at a few tens of minutes and hour scales at NIR \cite{Genzel-etal2003}, 43~GHz and 230~GHz \cite{Iwata-etal2017,Iwata-etal2020}. Even shorter periods, a few $10^3$ sec timescales are seen in X-ray flares of Sgr~A$^*$ \cite{Haggard-etal2019}. All these observations suggest the possibility of the presence of low angular momentum flow around Sgr~A$^*$. Moreover, recent observations of 230~GHz flux and the unresolved linear polarization signature of Sgr A$^*$ roughly agree with wind-fed low-angular momentum flow models, e.g., \cite{Murchikova-etal2022,Ressler-etal2023}. If it is true, we predict the observation of flickering in the liner polarisation pattern. 
More recently, in 2023 and 2024, JWST made continuous observations of Sgr~A$^*$ at $2.1$ and $4.8\,\mu$m wavelengths using NIRCam for $\sim48$ hours across seven epochs \cite{Yusef-Zadeh-etal2025}. These data indicate that Sgr~A$^*$ flux fluctuates constantly. They present the evidence of sub-minute (i.e., $10s$ of mHz), horizon-scale time variability of Sgr~A$^*$. From our analysis, this strongly suggests the presence of low angular momentum flow in the near-horizon region of Sgr A$^*$.

In the end, we would like to argue that the previous models \cite{Titarchuk-Fiorito2004,Cabanac-etal2010,Das-etal2014,Tagger-Pellat1999,Varniere-etal2002,Stella-Vietri1998,Schnittman-etal2006,Ingram-eta2009} for QPOs, none of them can explain such a range of QPOs (i.e., 10s mHz for $10^7M_\odot$ SMBHs). Our study shows a strong observational prediction of the evidence of cHz QPOs for low angular momentum flows around SMBHs. We should note that the choice of angular momentum to have perfect oscillations is very fine-tuned. Therefore, observations of them may be very difficult. Since these reported oscillations happen very close to the black hole event horizon, they may contain distinct signatures of spacetime around the black hole. Accordingly, these oscillations could be a potential tool to investigate physics under a strong gravity regime in the future.

\begin{acknowledgments}
This research is supported by the National Key R\&D Program of China (No. 2023YFE0101200), the National Natural Science Foundation of China (Grant No. 12273022), and the Shanghai Municipality orientation program of Basic Research for International Scientists (grant no. 22JC1410600). IKD acknowledges the TDLI postdoctoral fellowship for financial support. The simulations were performed on the TDLI-Astro cluster in Tsung-Dao Lee Institute, Pi2.0, and Siyuan Mark-I clusters in the High-Performance Computing Center at Shanghai Jiao Tong University.
This work has made use of NASA's Astrophysics Data System (ADS) We thank the referee for their insightful feedback, which improved the manuscript.
\end{acknowledgments}

\appendix

\section{:Long-term flux profiles}\label{Appendix-A}
%************************Figure s1******************************
\begin{figure}[h]
    \centering
    \includegraphics[width=0.5\textwidth]{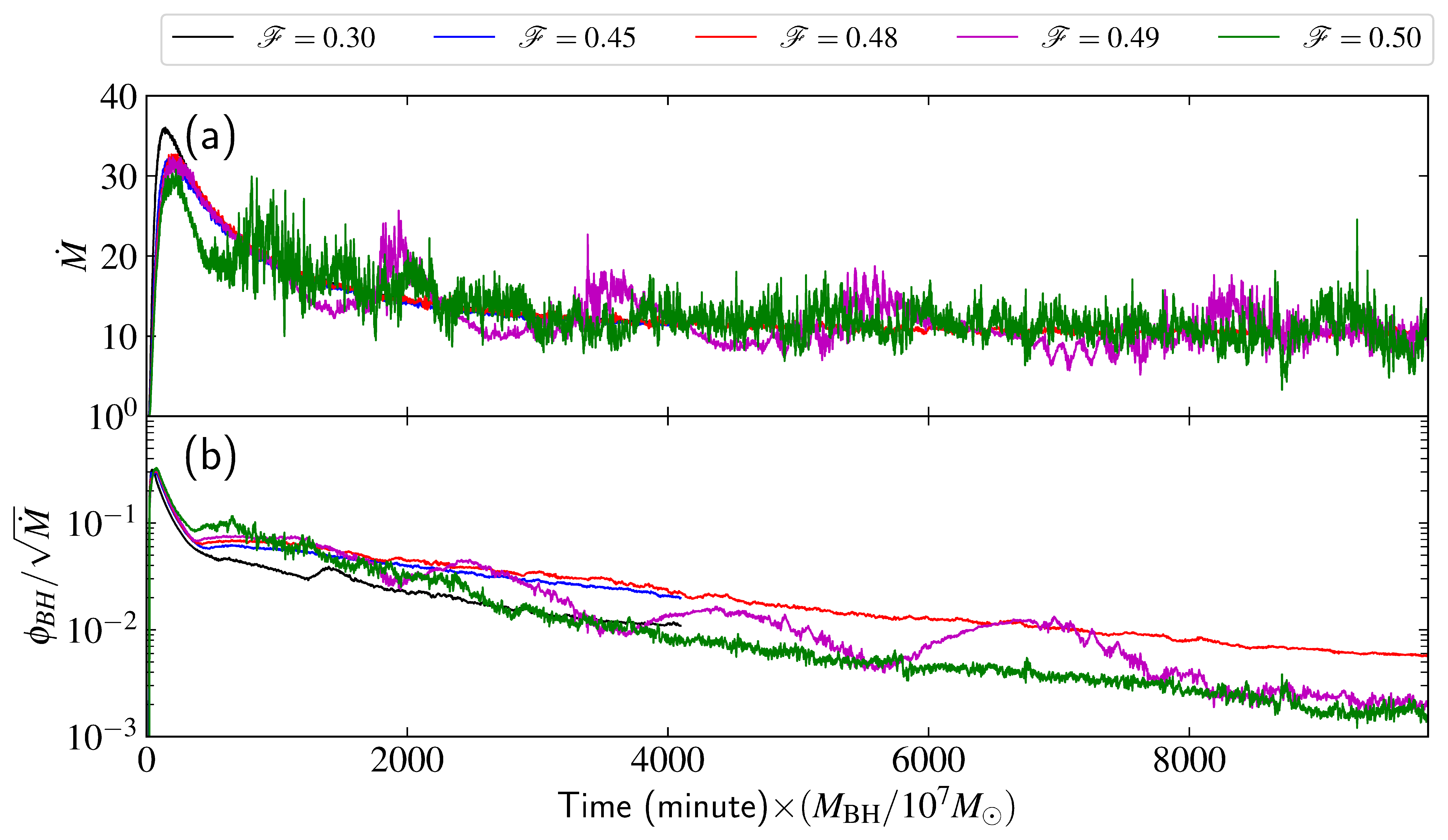}
    \caption{Evolution of mass accretion rate and normalized magnetic flux. Mass accretion rate ($\dot{M}$, panel a) and normalized magnetic flux ($\phi_{\rm BH}/\sqrt{\dot{M}}$, panel b) calculated at the horizon for different models with different ${\cal F}$.}
    \label{fig-s1}
\end{figure}
%***************************************************************
In Fig.~\ref{fig-s1}, we show long-term evolution of accretion rate ($\dot{M}$, in code units) and normalized magnetic flux ($\phi_{\rm BH}/\sqrt{\dot{M}}$) for our simulation models with different ${\cal F}$. Initially, the low angular momentum flow falls onto the black hole with very high mass flux, which is higher for the lower value of ${\cal F}$. Later, all the simulation models show quasi-steady state values after simulation time $t>3\,000\,t_g$ or $t>2000\,M_{\rm BH}/10^7M_\odot$ minutes. The lower angular momentum case does not show much variability in the value of the accretion rate. With the increase in angular momentum of accretion flows, variability in the accretion rate profile increases. We reported similar properties in our earlier study (for detail, check Refs. \cite{Dihingia-etal2024}). On the contrary, we do not see the reaching of a quasi-steady state in the normalized magnetic flux profiles. Due to the shorter infalling timescale, MRI cannot grow effectively for low-angular momentum flows. Accordingly, we observe a decaying nature in the magnetic flux profiles. This feature may be different if we increase the magnetic field strength or consider a very high resolution when we resolve small-scale MRI wavelength. That kind of study is computationally very expensive, but we plan to do it in the future. Nonetheless, with higher resolutions, the angular momentum ranges are expected to sift to the lower side to obtain similar results. In the next section \ref{Appendix-C}, we show the comparison of results in two different resolutions.

\section{Altitude effects}\label{Appendix-B}
%************************Figure s2******************************
\begin{figure}
\centering
\includegraphics[width=0.95\textwidth]{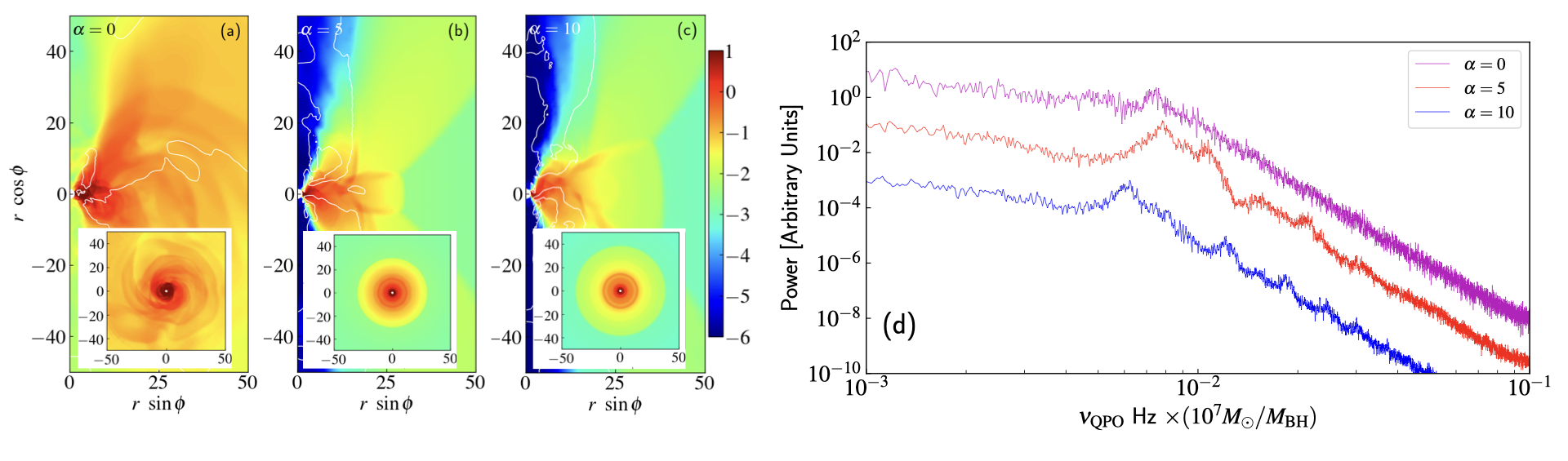}
\caption{Density distribution and PDS of accretion rate for different values of $\alpha$. The density distribution is presented at simulation time $t=5000\,t_g$. The simulations are used ${\cal F}=0.5$ but $\alpha$ value is changed as $0$, $5$, and $10$.
The white contour corresponds to $u^r=0$. In the inset, the same is shown on the equatorial plan.}
\label{Fig-s2}
\end{figure}
%***************************************************************

The initial density distribution of the accretion flow is given by $\rho (r,\theta)=\rho_{\rm FM}(r,\theta) \exp(-\alpha^2 \cos\theta^2)$, where any non zero value of $\alpha$ will flatten the initial density distribution. Accordingly, with an increase in $\alpha$, the inflow of matter is confined closer to the equatorial plane. When matter is accreting from very high altitudes ($\alpha=0$), we observe a turbulent accretion flow close to the black hole. In this case, relatively high-density matter occupies the region near the rotation axis. We do not see clear bipolar outflows. Outflow happens towards the equatorial plane. However, with the increase in $\alpha$ (panel 
Fig.~\ref{Fig-s2}b, we observe the formation of bipolar outflows around the rotation axis clearly. Panels of Fig.~\ref{Fig-s2}b and ~\ref{Fig-s2}c also show a sharp density jump in the distribution which corresponds to standing shocks. With the reduction of $\alpha$, standing shocks form far from the black hole. For these models, we find the shock locations to be more or less steady with some oscillations. In Fig.~\ref{Fig-s2}(d), we can see clear peaks in the power density spectra when accretion is happening closer to the equatorial plane ($\alpha \gg 1$). However, as the altitude of accretion increases ($\alpha=0$), we do not observe clear peaks in the PDS. Note that for $\alpha=0$, the accretion flows close to the black hole are more turbulent as compared to $\alpha>0$ (see panel of Fig.~\ref{Fig-s2}a). However, the turbulence is random, and accordingly, it does not appear in the PDS. 

\section{Resolution test}\label{Appendix-C}

\begin{figure}
\centering
\includegraphics[width=0.95\textwidth]{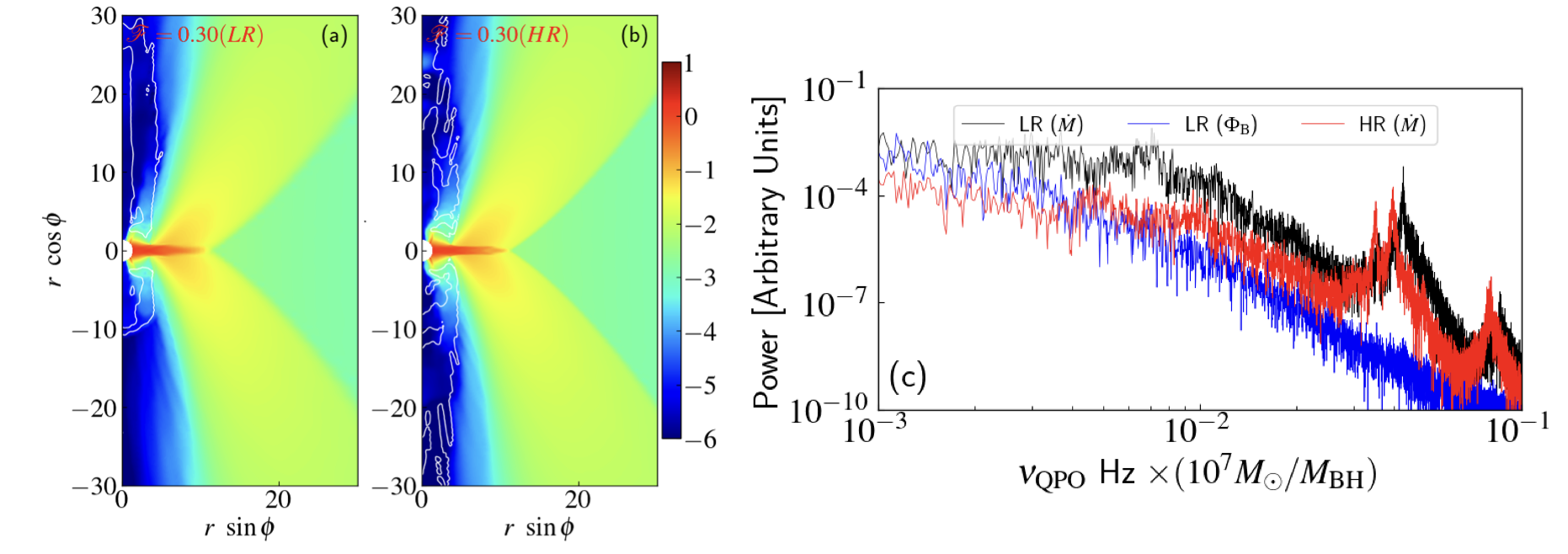}
\caption{Density distribution and PDS for two different resolution cases. It shows the density distribution for the case ${\cal F}=0.30$ with $a=0.94$ at simulation time $t=5000\,t_g$ (panel (a) and (b)) and the corresponding PDS of mass accretion and normalized magnetic flux for these models in panel (c). The white contour in panels (a) and (b) corresponds to $u^r=0$.}
\label{Fig-s3}
\end{figure}

Spatial resolution plays a crucial role in GRMHD simulations. Thus it is important to check if simulation results are consistent with different resolutions. To check it, we show the density distribution for ${\cal F}=0.30$ in panels Fig.~\ref{Fig-s3}(a) and ~\ref{Fig-s3}(b) at simulation time $t=5000\,t_g$. The resolutions are different but the simulation time is the same. In the low-resolution (LR) case, we have $256\times80\times64$ grids throughout the simulation box. For the high-resolution (HR) case, we resolve the region $\pm65^\circ$ from the equatorial plane and $r\le100\,r_g$ with $512\times160\times128$, and the outside of these regions is the same as the low-resolution case. In both panels (a and b), we see quite similar structures with the development of PS close to the black hole. Therefore, we anticipate that the resolution does not change our qualitative results. To check 
the results more quantitatively, we plot PDS of accretion rates at the event horizon (with $t>3000\,t_g$) for these two cases in the panel of Fig.~\ref{Fig-s3}(c). For comparison, we also show the PDS corresponding to the magnetic flux. We find similar PDS with prominent peaks in both cases. However, as the resolution increases, the value of peak frequency shifts to a lower value. In the high-resolution case, MRI is resolved better than in the low-resolution case. This results in slightly higher magnetic pressure in the high-resolution case. Accordingly, we observe a slightly higher outflow region and a slightly bigger PS in the high-resolution case. This increases the oscillation time scale of it, resulting in the shift of the frequency peak towards the lower side.
By comparing black and blue PDS, we do not find any peak in the PDS of magnetic flux. This suggests that the magnetic turbulence is not responsible for the observed peaks in PDS, but rather that angular momentum (or the in-fall properties) plays a decisive role in the overall flow properties.

%************************Figure s3******************************

\bibliography{sample631}{}
\bibliographystyle{aasjournal}



\end{document}